%% file: sample631.tex
\documentclass{aastex631}
\begin{document}

\title{Optimization of performance of the KM2A full array using the Crab Nebula}
\input{name}
\shortauthors{LHAASO Collaboration}
\correspondingauthor{X.Q.DONG, C.LI, H.H.HE}
\email{dongxuqiang@ihep.ac.cn, licong@ihep.ac.cn, hhh@ihep.ac.cn}

\begin{abstract}

The full array of the Large High Altitude Air Shower Observatory (LHAASO) has been in operation since July 2021. For its kilometer-square array (KM2A), we have optimized the selection criteria for very high and ultra-high energy $\gamma$-rays, using the data collected from August 2021 to August 2022, resulting in an improvement on significance of about 15$\%$ compared with previous cuts. With the implementation of these new selection criteria, the angular resolution is also significantly improved by approximately 10$\%$ at tens of TeV. Other aspects of the full KM2A array performance, such as the pointing error are also calibrated using the Crab Nebula. The resulting energy spectrum of the Crab Nebula in the energy range of 10-1000 TeV can be well fitted by a log-parabola model, which is consistent with the previous results from LHAASO and other experiments.

\end{abstract}


\section{ Introduction}  \label{sec:Intro}

The Crab Nebula is one of the very few celestial bodies corresponding to a recorded historical supernova explosion. It is powered by a spinning pulsar with a 33 ms period and a fluctuating magnetized relativistic pulsar wind. It is well studied in almost all wavelength bands from radio to gamma rays. The photons with energy just below 1 GeV are produced by synchrotron radiation, and the higher energy signal is dominated by inverse Compton scattering \citep{2004ApJ}. Meanwhile, the Crab Nebula is one of the most energetic sources in the TeV $\gamma$-ray energy band.

In 1989, very high energy (VHE) $\gamma$-ray emission from the Crab Nebula was first discovered by the Whipple Collaboration \citep{crab}. Since then, it has been detected by both Cherenkov telescopes \citep{2004ApJ} \citep{Aharonian_2006} and air shower arrays \citep{Bartoli} \citep{PhysRevD} 
 \citep{2019ApJ} \citep{EAS}. At the lower energy band, MAGIC has measured the spectrum down to 77 GeV \citep{Aleksi2014}. The highest energy range was explored by air shower arrays. HAWC \citep{Abeysekara_2019} and Tibet AS$\gamma$ \citep{ASr100TeV} detected $\gamma$-ray signals around 100 TeV from the Crab Nebula. The result from LHAASO has extended the spectrum beyond 1 PeV \citep{LHAASO:science} \citep{LHAASO:Nature}, which implies the presence of a PeV electron accelerator. Considering the stable flux and well consistent measurements from different experiments, it is commonly used as a ``standard candle" to check detector performance, including pointing accuracy, angular resolution, background rejection power, flux determination, etc.

With the start of operation of the KM2A full array since 2021, we have restudied the performance of the array and optimized the $\gamma$-ray selection criteria, resulting in a significant improvement in sensitivity. After describing the KM2A full array in section 2, we present the optimization of the KM2A full array in section 3. We present the observation of Crab Nebula in section 4. In the last section we discuss our results and draw conclusions.

\section{LHAASO and KM2A full array} \label{s:sec1}
LHAASO (100.01$^\circ$ E, 29.35$^\circ$ N) is a large hybrid extensive air shower (EAS) array located at Haizi Mountain in Daocheng, Sichuan Province, China, which was fully constructed in July 2021. It consists of three sub-arrays \citep{hhh}: the Kilometer-Squared Array (KM2A) for $\gamma$-ray astronomy above 10 TeV and cosmic ray physics; Water Cherenkov Detector Array (WCDA) for $\gamma$-ray astronomy above a few hundreds of GeV; and 18 Wide field-of-view air Cherenkov telescopes (WFCTA) for cosmic ray physics from 10 TeV to 1 EeV. KM2A has a field-of-view (FOV) of ~2 sr and covers 60\% of the sky. A study of the performance of the KM2A half array has been published \citep{2021Observation}.

\begin{figure}[htbp]
\centering
\includegraphics[width=0.4\textwidth]{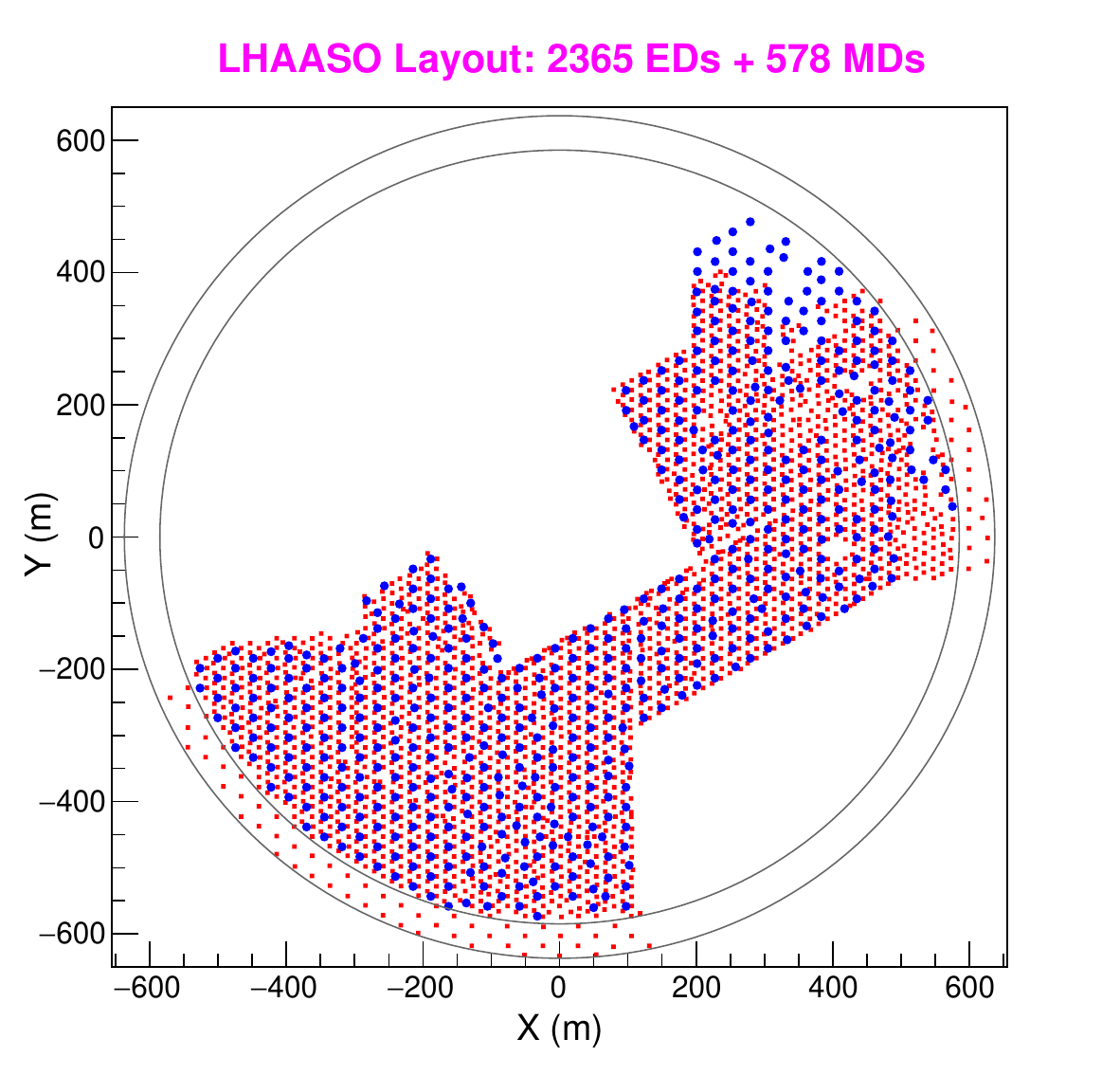}
\includegraphics[width=0.4\textwidth]{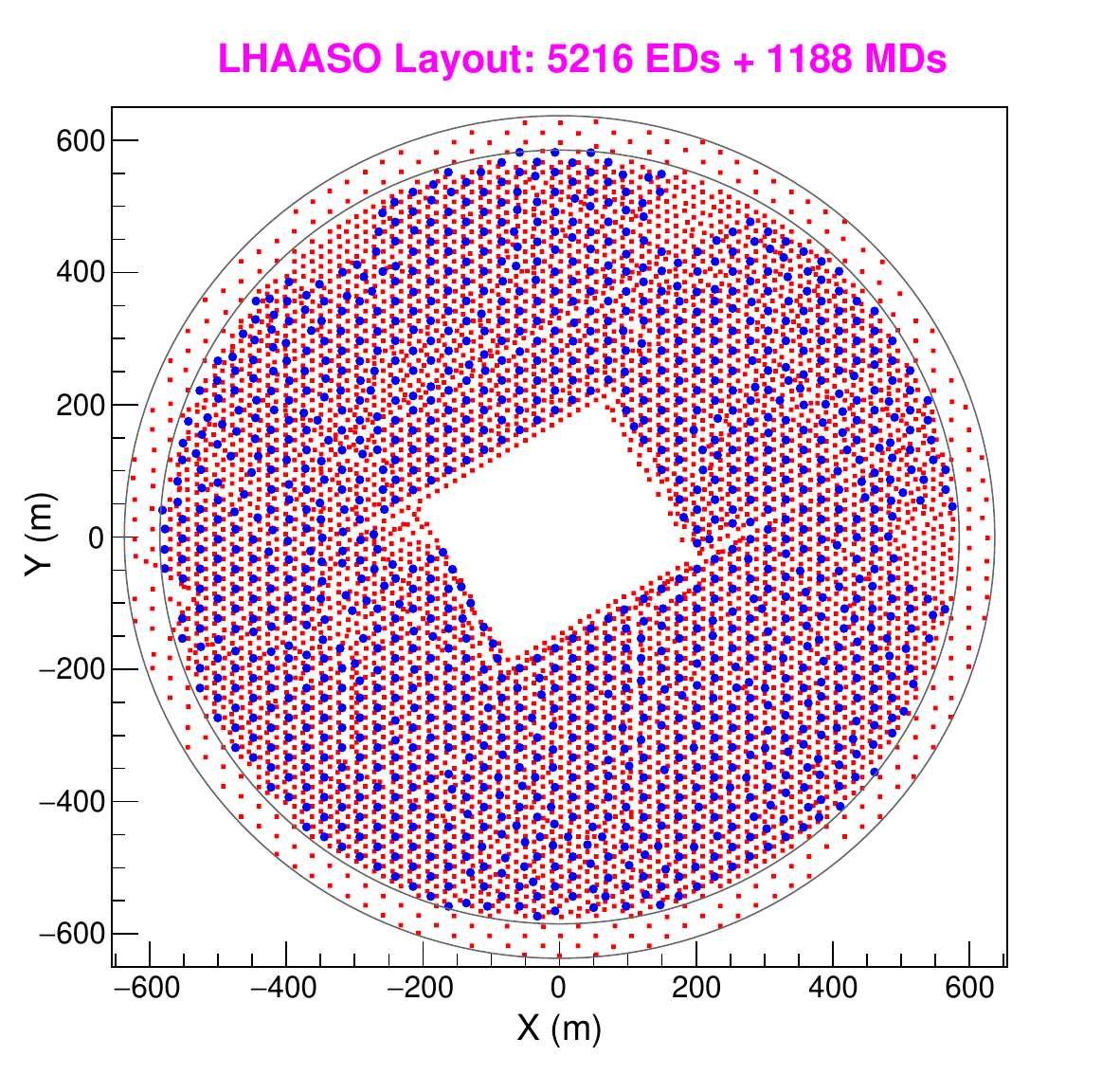}
\caption{LHAASO layout: the red squares represent the electromagnetic detectors (EDs) and the blue circles represent the muon detectors (MDs). The left panel shows the KM2A half array finished in 2019, which consists of 2365 EDs and 578 MDs; the right panel shows the KM2A full array finished in 2021, which consists of 5216 EDs and 1188 MDs. The central white rectangle in the right panel indicates the LHAASO-WCDA array region.}
\label{fig:my_label1}
\end{figure}

The full array of KM2A started operation in 2021. The layouts of the KM2A half array and full array are shown in Figure \ref{fig:my_label1}. Compared with the half array, not only the number of detectors increased, but the detectors are more uniformly distributed as well. Thus the performance should be different for the two configurations. In principle, the sensitivity of the full array can be increased with more suitable data selection criteria.

The event reconstruction algorithm was developed and applied to the data of the half array \citep{2021Observation}. The lateral distribution of shower secondary particles is fitted by a modified NKG function and the density at a fixed distance of 50 m from shower axis ($\rho_{50}$) is used to estimate the primary energy. The primary direction is reconstructed by fitting the relative arrival times of shower particles to a conical plane \citep{HAWC_LDF}. The pointing accuracy of 1/2 KM2A for $\gamma$-ray events is estimated to be better than 0.1$^\circ$ and its angular resolution is less than 0.3$^\circ$ above 100 TeV \citep{2021Observation}. 

In this study, the Monte Carlo (MC) data for $\gamma$-ray showers were obtained from CORSIKA \citep{Heck1998CORSIKAAM} and G4KM2A \citep{Jin_2020} \citep{AGOSTINELLI2003250}, as previously described by \citep{2021Observation}. The same event reconstruction algorithm is applied to the full array data.

\section{Optimization of gamma-ray selection criteria} \label{s:sec2}
In this paper, we aim to improve the performance of the full array for $\gamma$-ray source detection by optimizing the data selection criteria. The data selection criteria in 
\cite{2021Observation} are: (1) the shower core is located in the area of the KM2A half array shown in the left panel of Figure \ref{fig:my_label1}; (2) the reconstructed zenith angle is less than $50^{\circ}$; (3) the number of particles detected within 40 m from the shower core is larger than that within 40-100 m; (4) the number of EDs and the number of particles for the reconstruction are both greater than 10; and (5) the shower age is between 0.6 and 2.4. The cut on the value of the $\gamma$/hadron discrimination parameter varies with different energy bands. In this paper, the data selection criteria in \citep{2021Observation} are called the old data selection criteria and here we derive new data selection criteria after optimization.  

\subsection{\texorpdfstring{$\mathcal\gamma/$}/hadron discrimination optimization} \label{s:sec2.2}
The low proportion of $\gamma$-rays in cosmic rays makes it challenging to distinguish them, which is crucial for ground-based experiments. Various experiments employ different methods to address this issue. The Cherenkov telescopes use the shape of the image to discriminate the two kinds of signals. Water Cherenkov detectors, such as HAWC \citep{HAWC_LDF} \citep{PINC} and LHAASO-WCDA \citep{compactness}, take advantage of the different lateral distributions of showers.

When passing through the Earth's atmosphere, a $\gamma$-ray will interact with an atomic nucleus in the atmosphere and generate an electromagnetic shower, which is muon-poor, but a background cosmic ray ion will generate a hadronic shower, which is muon-rich. Therefore, the ratio between the measured number of muons and electrons is utilized by LHAASO-KM2A to discriminate $\gamma$-rays from cosmic ray ions. In this paper, the same parameter is used to discriminate between $\gamma$-rays and hadrons \citep{2021Observation}.

\begin{figure}[htbp]
\centering
\includegraphics[width=0.5\textwidth]{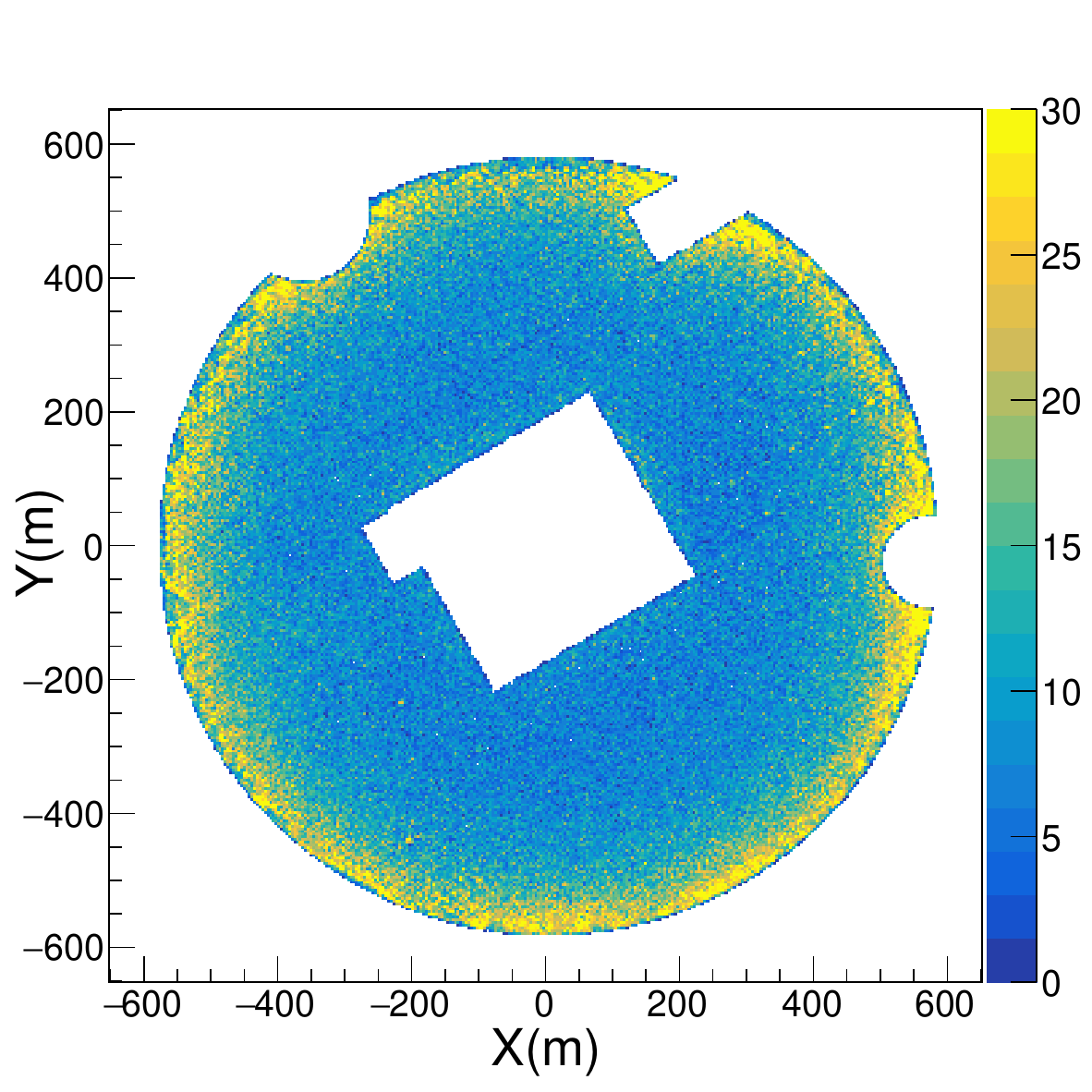}
\caption{The distribution of reconstructed core positions for cosmic-ray background events in KM2A using experimental data after applying the old data selection criteria. The color represents the number of cosmic-ray background events, with the maximum value set to 30. The regions of missing events at the edges of the circle are due to the absence of detectors at those positions.}
\label{fig:my_label3}
\end{figure}

\begin{figure}[htbp]
\centering
\includegraphics[width=0.5\textwidth]{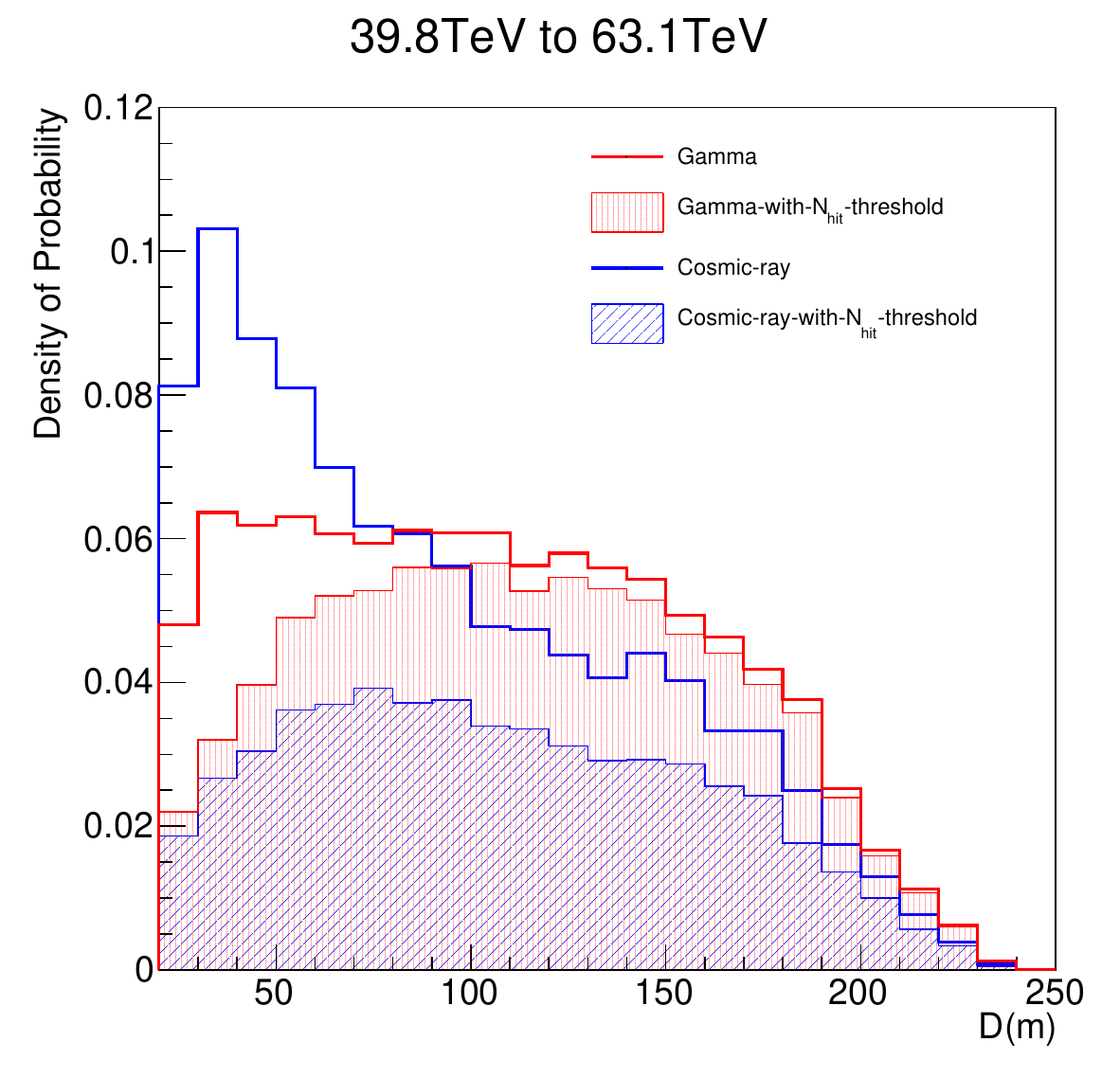}
\includegraphics[width=0.5\textwidth]{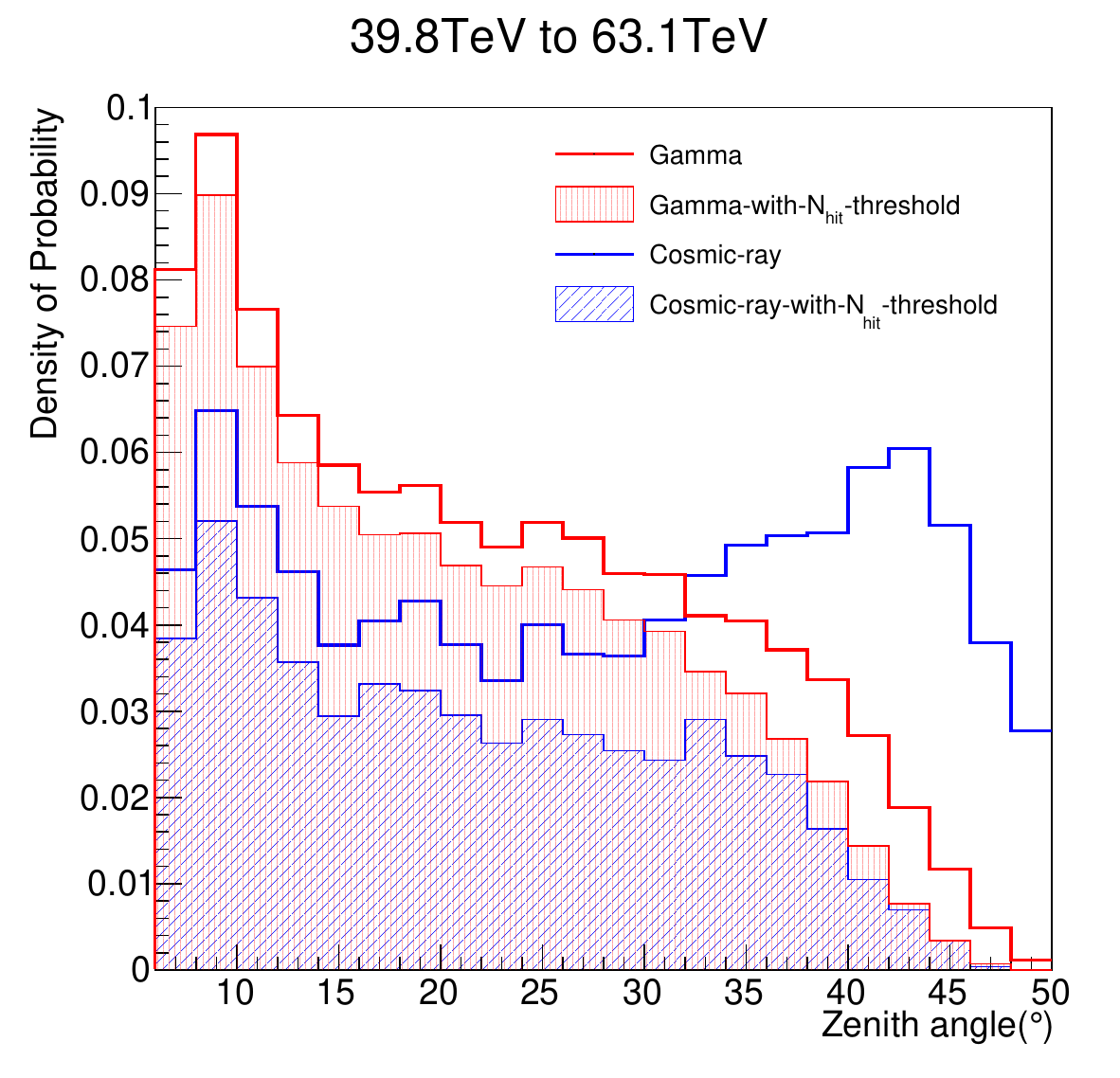}
\caption{The distributions of $D$ (distance from the shower core position to the edge of the detector, top) and zenith angle (bottom) before and after applying the threshold for $N_{hit}$. The red and blue histograms represent gamma-rays and cosmic-rays separately.}
\label{fig:my_label5}
\end{figure}

\begin{figure}[htbp]
\centering
\includegraphics[width=0.5\textwidth]{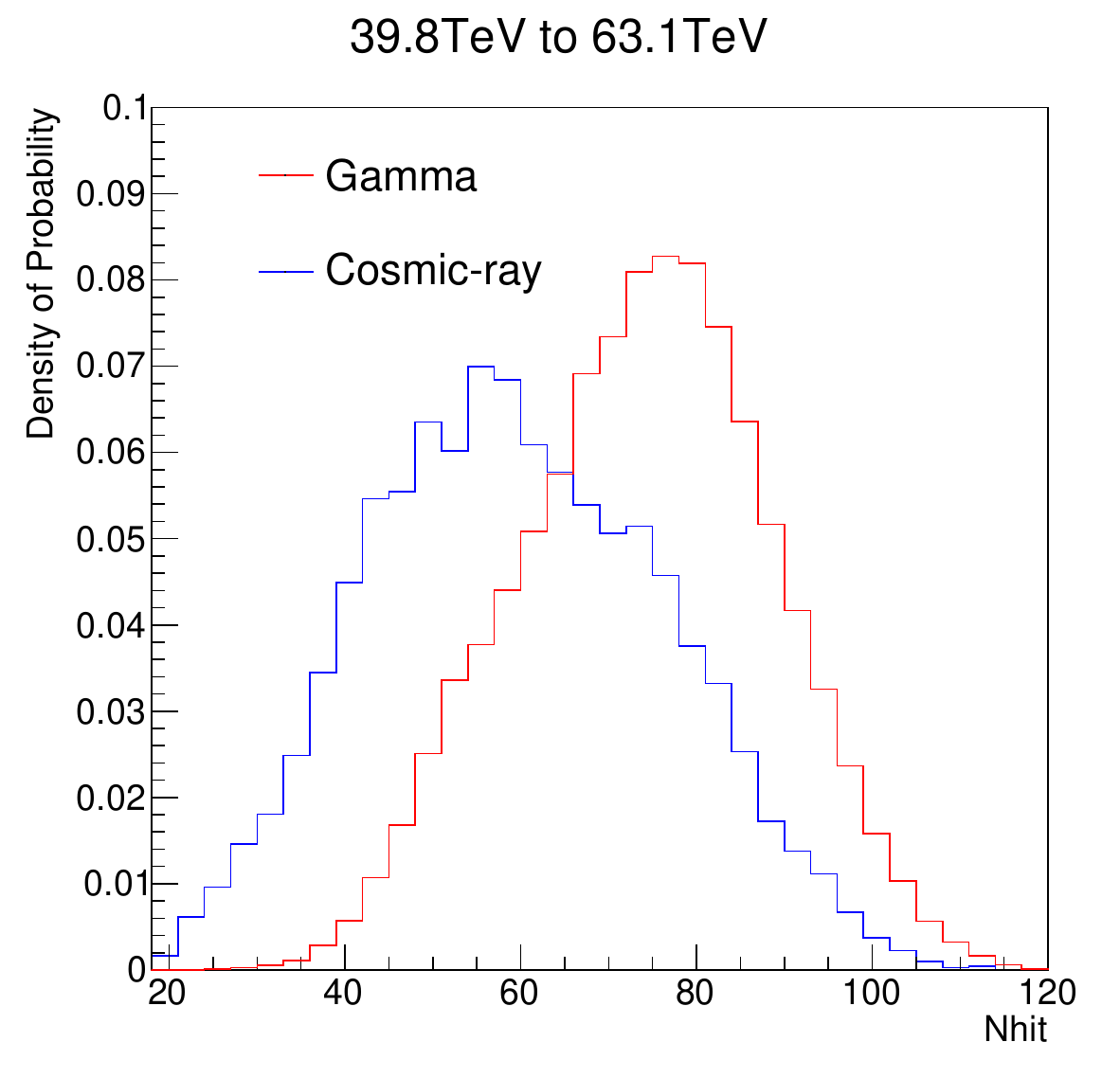}
\caption{The $N_{hit}$ distributions of simulated $\gamma$-ray (red) and observed cosmic-ray events after $\gamma$/hadron discrimination (blue) in reconstructed energy bin from 39.8TeV to 63.1TeV.}
\label{fig:my_label4}
\end{figure}

Figure \ref{fig:my_label3} shows the reconstructed core distribution of the cosmic-ray events observed from August 2021 to August 2022 after applying the old data selection criteria. It is clear that the density is higher at the edge of the array, i.e, for a low distance $D$ from the edge. 
 As shown in Figure \ref{fig:my_label5}, there are also more cosmic-ray events at larger zenith angles, while the distributions in $D$ and zenith angle are different for simulated gamma-ray events, which indicates a weaker rejection power at these regions. In fact, the rejection power is directly related to $N_{hit}$ (the number of fired detectors), and cosmic-ray events with fewer $N_{hit}$ have a higher possibility to be misidentified as a photon-like event. Thus the cosmic-ray events are mainly concentrated at the edge of the array or at large zenith angles, where the rejection power is weaker. The distributions of $N_{hit}$ for cosmic-ray events and $\gamma$-ray showers are shown in Figure \ref{fig:my_label4}. The data for $\gamma$-rays are obtained from simulation, and the data for the cosmic-ray background events are from experimental measurements considering much larger statistics than the simulation and with a negligible contribution from gamma rays. It is obvious that the $N_{hit}$ of cosmic-ray showers is usually less than that of $\gamma$-ray showers. Therefore,we can further reduce the background by applying a cut on $N_{hit}$. The numbers of cosmic-ray events near the edge of the array and at large zenith angles are great reduced with the implementation of a threshold for $N_{hit}$ and the distributions of $D$ and zenith angle for cosmic ray and gamma events are similar now (see Figure \ref{fig:my_label5}). 

Based on the information mentioned above, we set a $N_{hit}$ threshold for each energy bin below 100 TeV by maximizing the significance of the Crab Nebula as shown in Table \ref{table:1}. 

\begin{center}
\begin{table}[htbp]
\centering
\begin{tabular}{c|c|c|c|c|c}
\hline
$E_{rec}/\text{TeV}$& (10.0,15.8)& (15.8,25.1)& (25.1,39.8)& (39.8,63.1)& (63.1,100.0)\\
\hline
$N_{hit}$ &20&30&44&58&64\\
\hline
\end{tabular}
\caption{The optimized $N_{hit}$ threshold for each energy $E_{rec}$ bin under 100 TeV.}
\label{table:1}
\end{table}
\end{center}

\begin{figure}[htbp]
\centering
\includegraphics[width=0.6\textwidth]{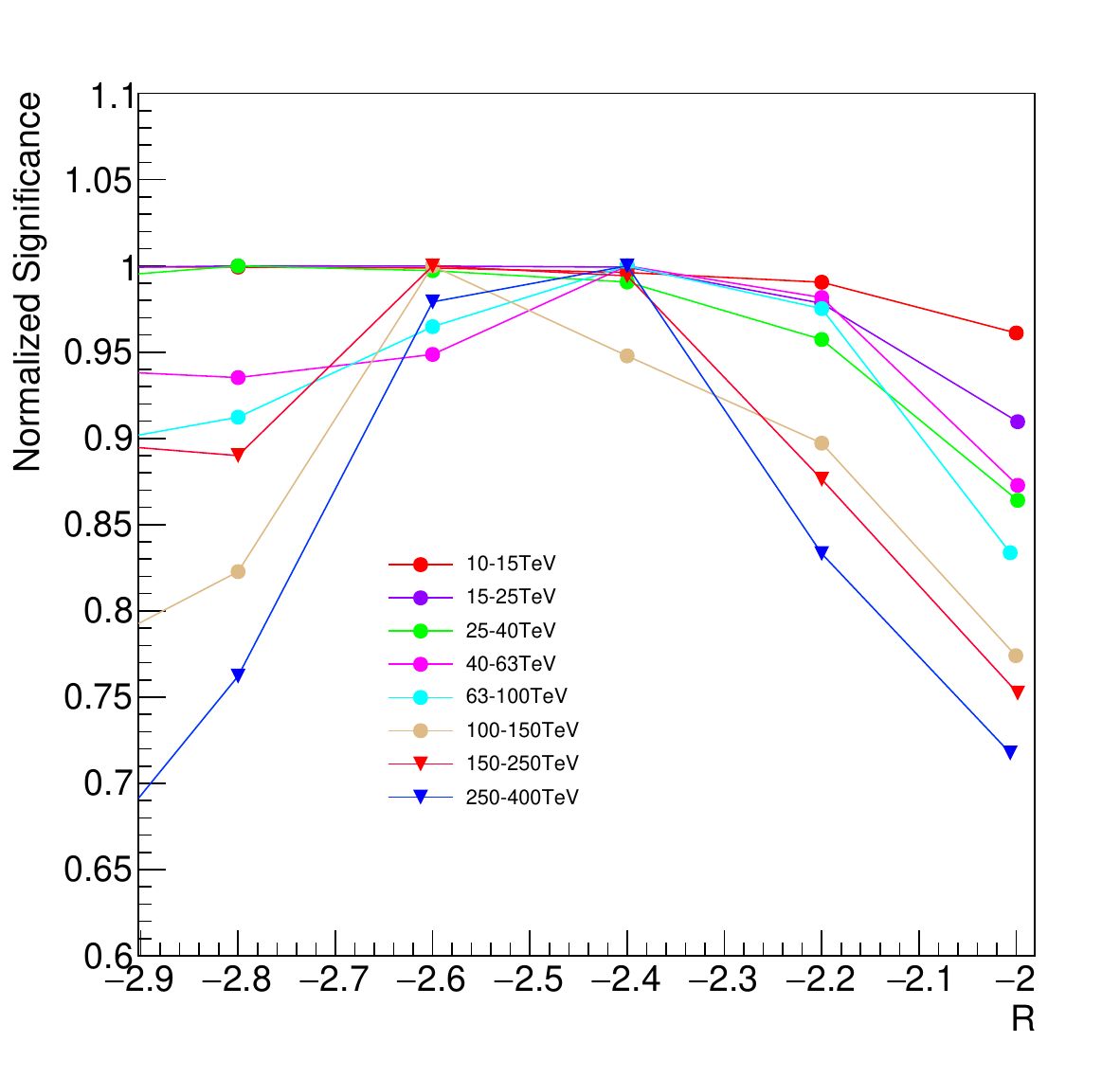}
\caption{The distribution of normalized significance of the Crab Nebula with different $R$ cut levels at each energy band. All datasets are normalized to their maximum values for better comparability. The normalized significance is calculated as the ratio of the significance to the maximum significance.}
\label{fig:my_label6}
\end{figure}

After the $N_{hit}$ cut is applied, it is found that the significance of the Crab Nebula reaches its maximum in each reconstructed energy bin with a consistent cut of $R$ around -2.4, as shown in Figure \ref{fig:my_label6}. This indicates that the $\gamma$/hadron selection criteria remains constant with respect to the shower energy, in contrast with the old criteria.

Finally, the following new data selection criteria were established:  (1) the shower core is located in the area of the KM2A full array shown in the right of Figure \ref{fig:my_label1}; (2) the zenith angle is less than $50^{\circ}$; (3) the number of particles for the reconstruction must be greater than 10; (4) for all reconstructed energy bins, the $R$ must be less than -2.4; and (5) for each energy bin, distinct thresholds for $N_{hit}$ were established. 

\subsection{Results of optimized \texorpdfstring{$\mathcal\gamma/$}/hadron discrimination} \label{s:sec2.4}

\begin{figure}[htbp]
\centering
\includegraphics[width=0.6\textwidth]{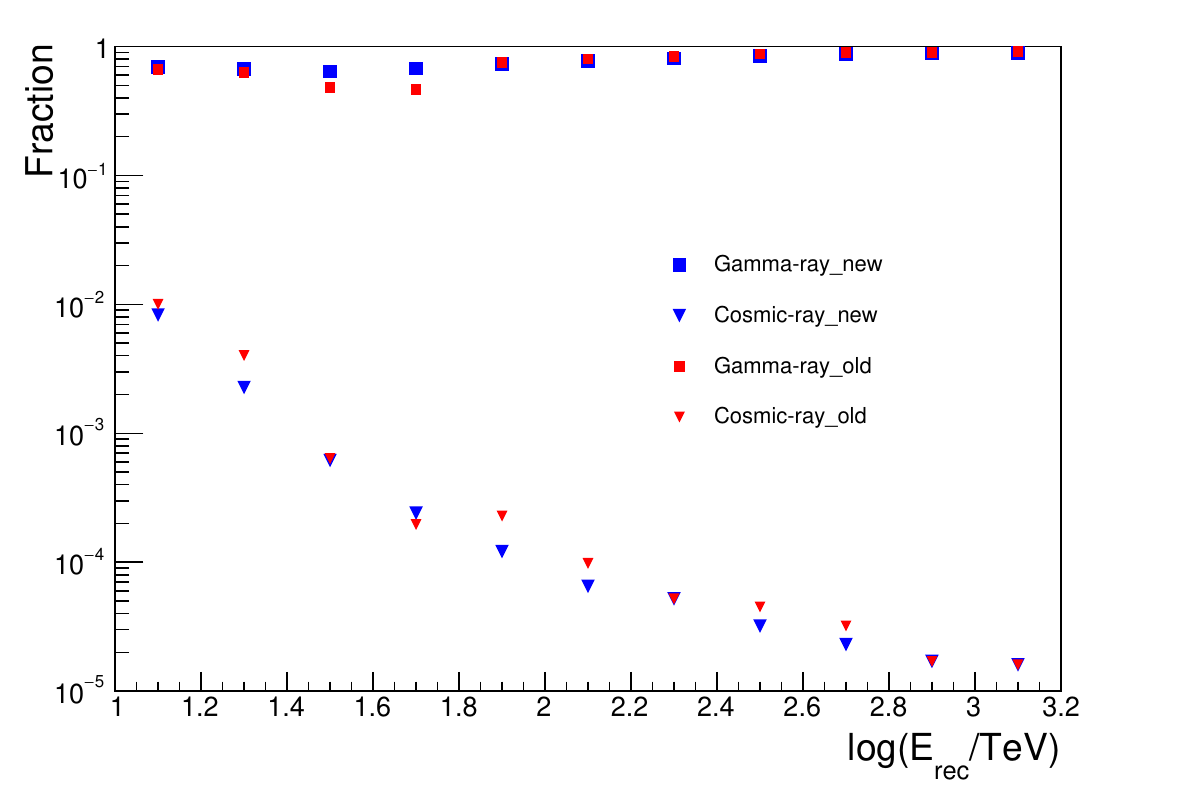}
\caption{The survival fraction of $\gamma$-ray events (represented by squares, as per simulation) and cosmic ray background events (represented by inverted triangles, as per observational data) varies with energy after applying $\gamma$/hadron discrimination cuts. A blue symbol represents the survival fraction of $\gamma$-rays and cosmic rays with the new data selection criteria, while the red symbol represents the fraction with the old data selection criteria. }
\label{fig:my_label7}
\end{figure}

Even though a large fraction of cosmic-ray events are rejected by the criteria above, especially using $R$, the muon cut, there are still residual events. Figure \ref{fig:my_label7} shows the survival fraction, defined as the ratio of the number of events after $\gamma$/hadron discrimination to the total number of $\gamma$-ray or cosmic-ray events. The fraction varies from $64\%$ to $89\%$ for $\gamma$-rays. The rejection power of cosmic-ray induced showers is better than $1.5\times10^4$ above 100 TeV, which is increased by about 5 times compared with the value for the half array using old selection criteria. What's more,

when applying these new data selection criteria, the evolution of the survival fraction with energy is smoother in comparison with the application of old cuts.

\begin{figure}[htbp]
\centering
\includegraphics[width=0.6\textwidth]{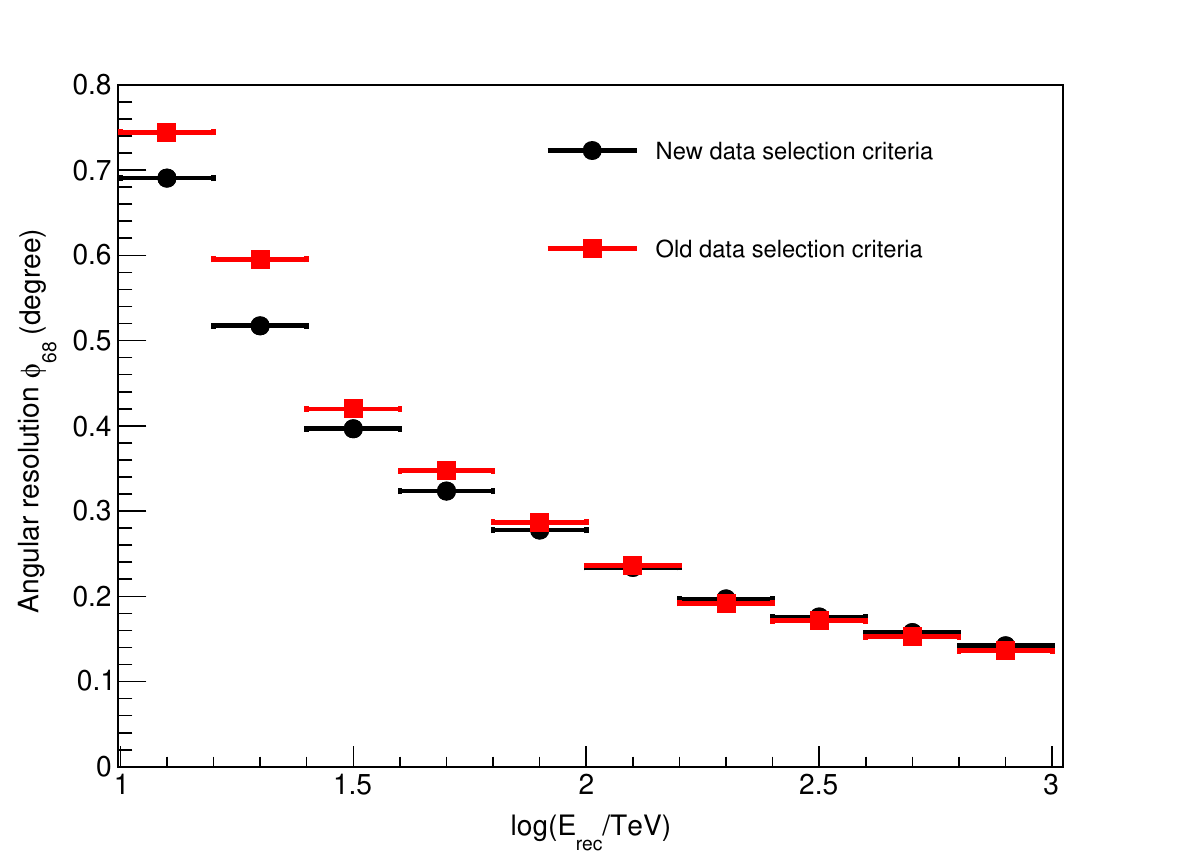}
\caption{Angular resolution of the KM2A full-array for simulated $\gamma$-ray showers with zenith angle less than 50$^{\circ}$. Black points show the angular resolution with the new data selection criteria, red squares are for the old data selection criteria.}
\label{fig:my_label8}
\end{figure}

With the $N_{hit}$ threshold set for each energy bin, not only the rejection power but also the angular resolution,  as shown in Figure \ref{fig:my_label8},  is improved especially at lower energies. The resolution (denoted as $\phi_{68}$, containing $68\%$ of the events) is now about $0.5^{\circ}$ at 20 TeV, whereas the previous value was $0.6^{\circ}$ using old cuts.  

\begin{figure}[htbp]
\centering
\includegraphics[width=0.6\textwidth]{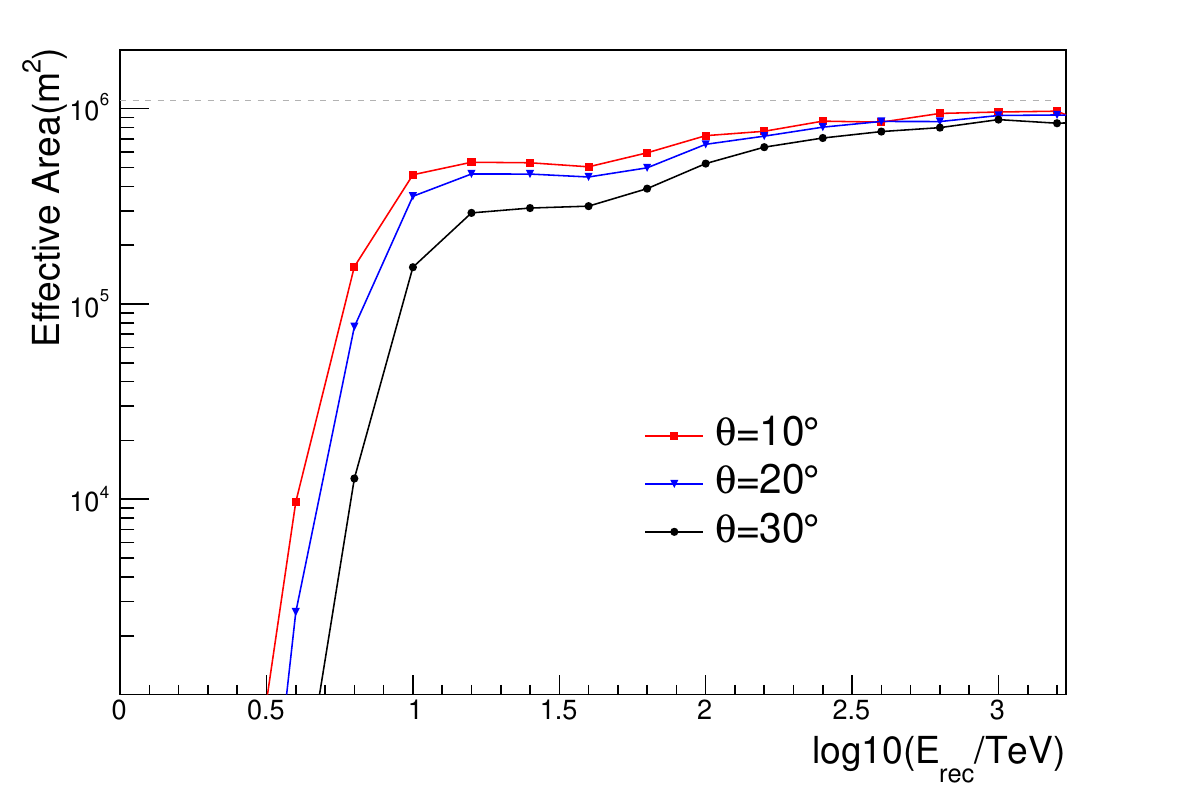}
\caption{Effective area of KM2A full-array for $\gamma$-ray showers at three zenith angle ranges after applying the data selection criteria. The error bars are too small to be seen. The gray dashed line represents the actual area of the detector.}
\label{fig:my_label9}
\end{figure}

The effective area of the KM2A array for gamma rays is also calculated using simulated events, which varies with zenith angle and energy. Figure \ref{fig:my_label9} shows the effective area of the KM2A full array at three zenith angles, $\theta=10^{\circ}, 20^{\circ}$ and $30^{\circ}$. The effective area increases with energy and reaches a constant value at an energy above 400 TeV. The effective area is more than $5\times10^{5}$ $\mathrm{m}^2$ above 10 TeV and $8\times10^{5}$ $\mathrm{m}^2$ above 100 TeV for a zenith angle of $10^{\circ}$.

\begin{figure}[htbp]
\centering
\includegraphics[width=0.6\textwidth]{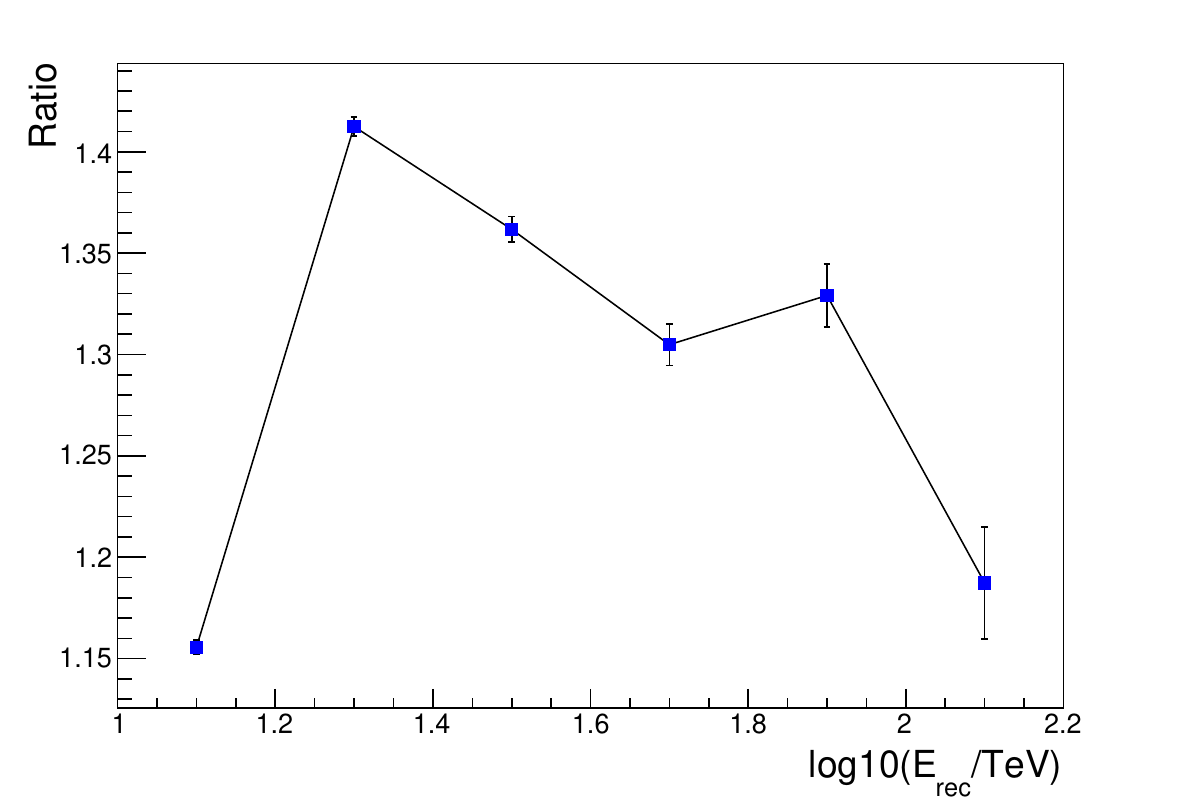}
\caption{Quality factor $ratio$=$Q_{new}/Q_{old}$ in different energy bins. The error bars represent the level of uncertainty in the measurement. }
\label{fig:my_label10}
\end{figure}

The ratio of quality factor, defined as $Q_{new}/Q_{old}$, is used to quantify the improvement in sensitivity. Here, $Q=\frac{\xi_\gamma}{\sqrt{\xi_p}}$, where $\xi_\gamma$ and $\xi_p$ represent the survival ratio of $\gamma$-rays and cosmic rays, respectively. The value of $\xi_\gamma$ is obtained from simulation and $\xi_p$ is derived from experimental data. Figure \ref{fig:my_label10} shows the values of the quality factor ratio at different energy bins. The ratio is approximately 1.41 at 20 TeV and around 1.18 at 125 TeV, which shows a clear improvement of performance at tens of TeV. 

\section{results for the Crab Nebula} \label{s:sec3}
 After optimizing the data selection criteria, we calibrated the performance of KM2A for $\gamma$-ray source detection using the Crab Nebula as a standard candle. The data used in this analysis were collected by the KM2A full-array from August 2021 to August 2022. The operational status of each detector is monitored in real time, and only detectors in normal condition are used in reconstruction. To ensure a stable array performance, the number of live EDs and MDs should be greater than 5200 and 1180, respectively. The total effective observation time was about 352 days. With a trigger rate of about 900 Hz, the number of events recorded by the KM2A full array was $7.6\times10^{10}$. The background estimation is performed using the direct integral method (DIM) \citep{DIM}, a widely adopted technique utilized by the ARGO-YBJ and HAWC experiments, as well as the KM2A half array \citep{2021Observation}.  The data selection criteria and the $\gamma$/hadron discrimination parameter were discussed in the previous section.

\begin{figure}[htbp]
\centering
\includegraphics[width=0.6\textwidth]{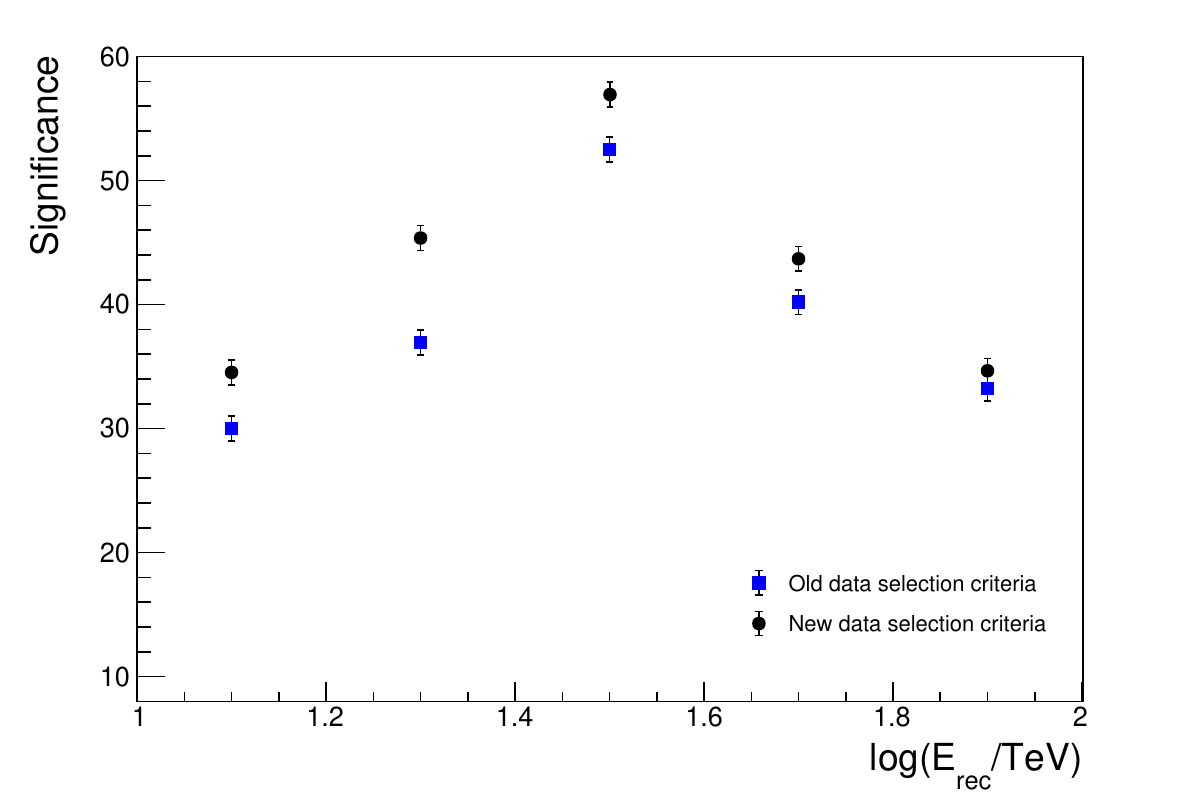}
\caption{Comparison of significance of detection of $\gamma$-rays from the Crab Nebula using two data selection criteria. The black dot represents the significance obtained using the new data selection criteria, while the blue square represents the significance obtained using the old data selection criteria. }
\label{fig:my_label12}
\end{figure}

Figure \ref{fig:my_label12} shows the significance of detection of $\gamma$-ray in each energy bin from the Crab Nebula using the two different data selection criteria.

It is obvious that the significance of detection $\gamma$-rays is increased with new data selection criteria. The differential significance is increased by up to $20\%$, and the integral significance is increased by approximately $15\%$. The Crab Nebula is observed at a significance of 51.0 $\sigma$ at 40-100 TeV and 23.1 $\sigma$ above 100 TeV.

\begin{figure}[htbp]
\centering
\includegraphics[width=0.6\textwidth]{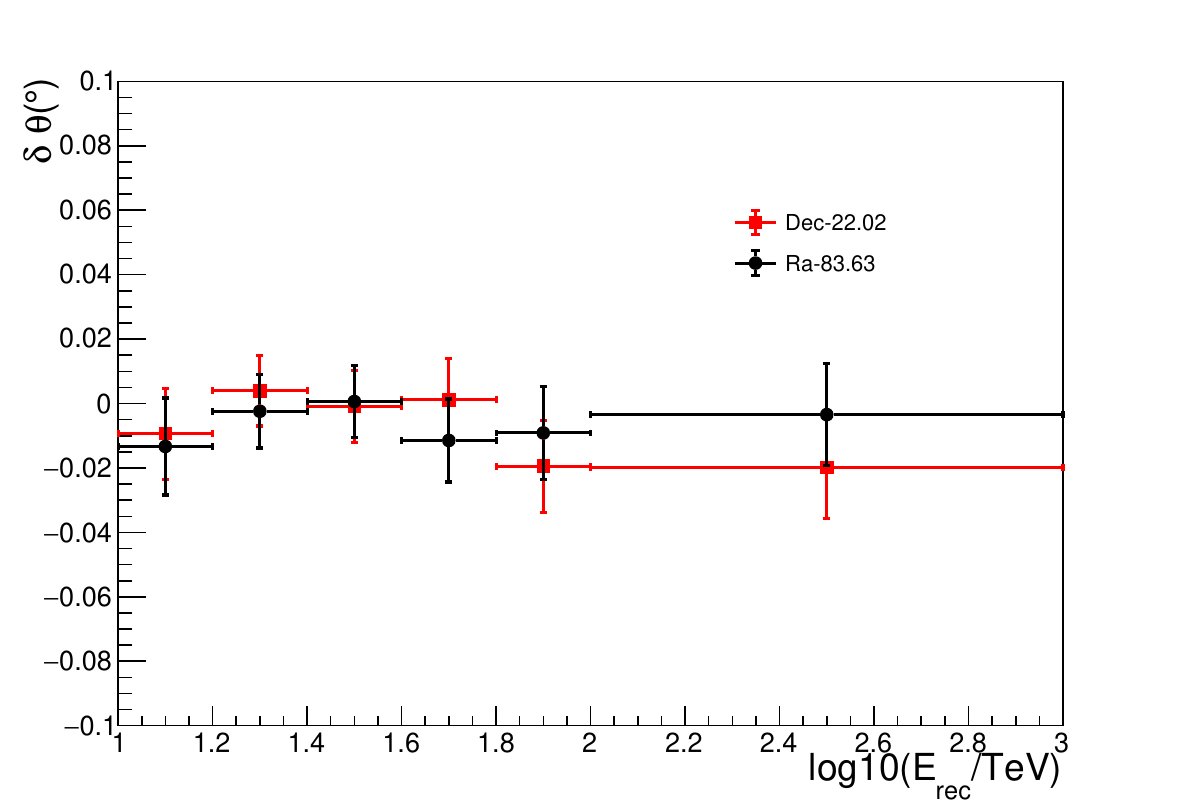}
\caption{Deviation of measured cntroid of $\gamma$-ray emission declination, right ascention from the Crab Nebula relation to the known position (22.02, 83.63) as a function of energy.}
\label{fig:my_label13}
\end{figure}

The position of the Crab Nebula $\gamma$-ray emission is fitted using a two-dimensional Gaussian function. The deviations in position, compared with the known declination and right ascension, obtained in different energy bins are illustrated in Figure \ref{fig:my_label13}. The pointing is consistent with the Crab Nebula's position within the 1 $\sigma$ statistical error. From the observations of the Crab Nebula, the pointing error of KM2A for $\gamma$-ray events is estimated to be less than $0.03^\circ$ even considering the statistical error.

\begin{figure}[htbp]
\centering
\includegraphics[width=0.5\textwidth]{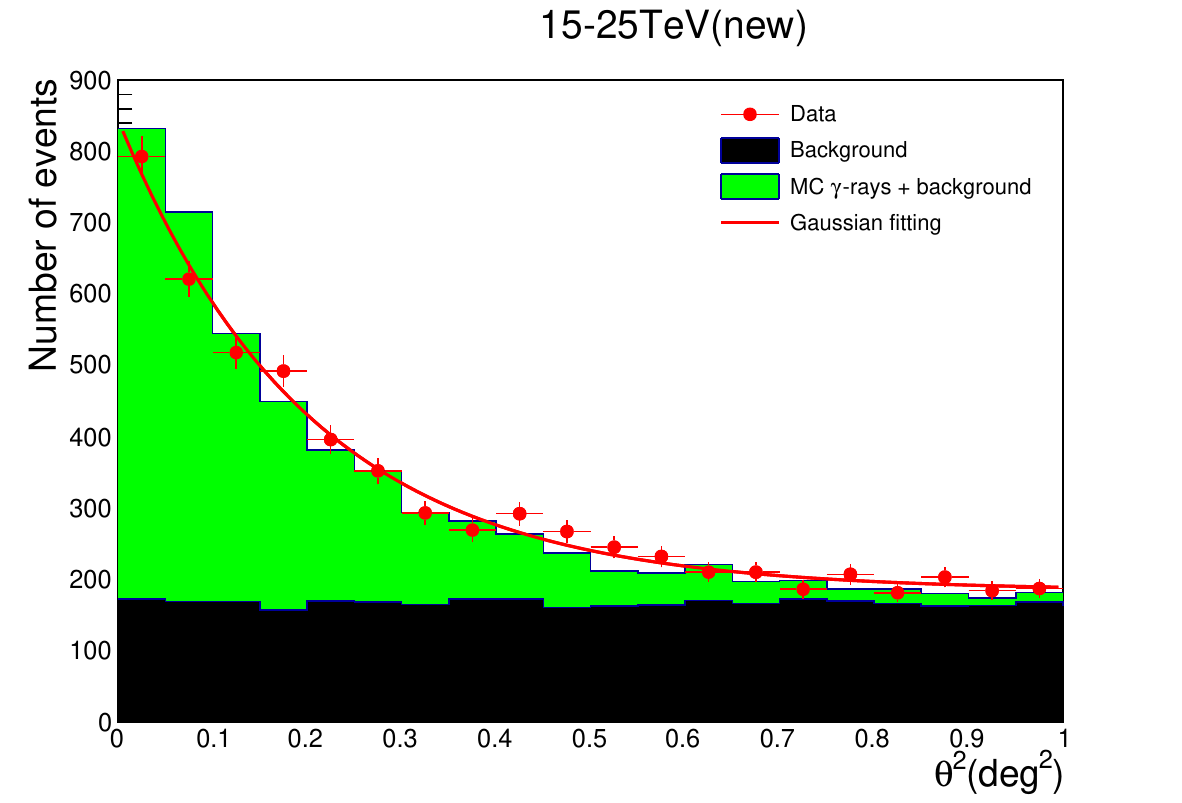}
\includegraphics[width=0.5\textwidth]{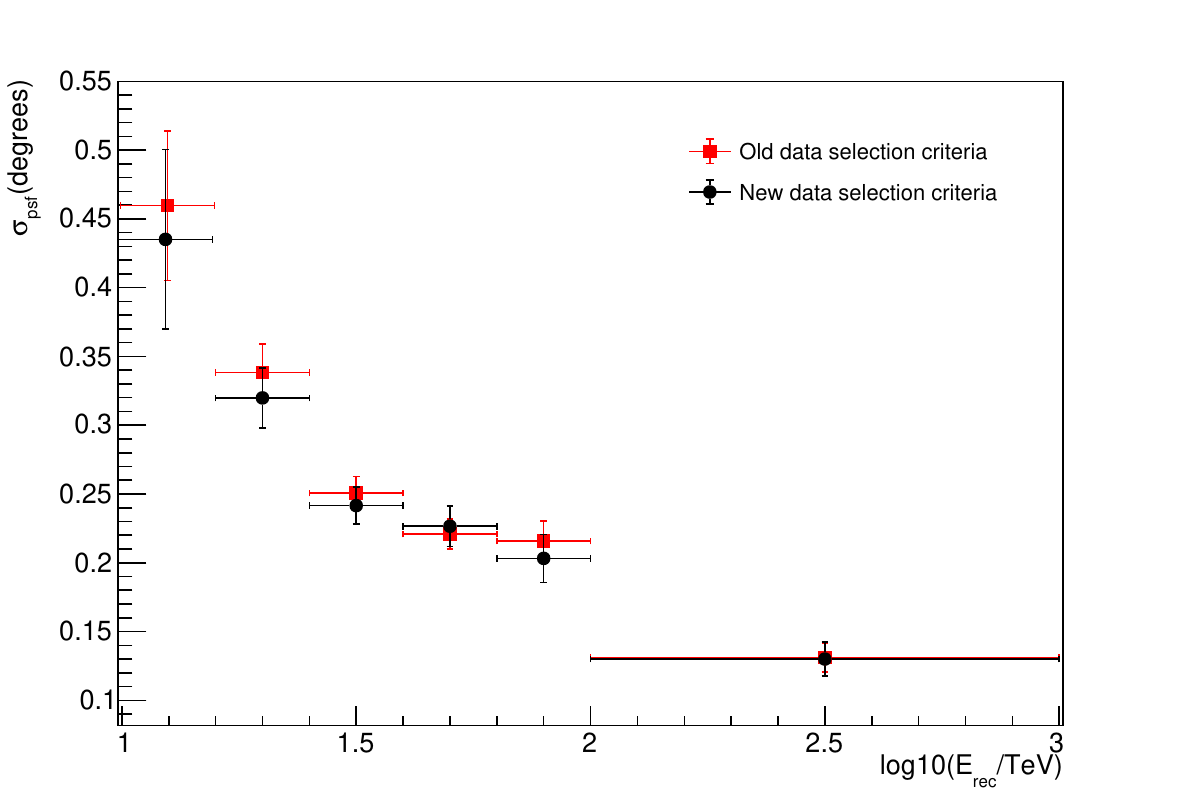}
\caption{Distribution of events as a function of the angle from the Crab Nebula direction for both experimental data and MC simulation. The top graph was obtained with the new data selection criteria. The bottom graph shows the angular resolution in different energy bins. The red squares shows the angular resolution from the old data selection criteria, and the black dots are from the new data selection criteria.}
\label{fig:my_label14}
\end{figure}

Figure \ref{fig:my_label14} shows the angular resolution obtained from observation of the Crab Nebula at different energy bins after the two $\gamma$/hadron discrimination cuts. The $\sigma_{psf}$ is obtained by fitting the angular distribution with a Gaussian function, which can be used to calibrate the performance of KM2A. Although the Crab Nebula $\gamma$-ray emission is slightly extended \citep{2020NatAs}, it is negligible compared with the  Pointing Spread Function (PSF) of KM2A. It is clear that the angular resolution is improved with these new selection criteria. The distribution of events as a function of angular distance from the Crab Nebula direction shows good consistency between simulated and observed data.

The energy spectrum is obtained by a 3D likelihood method \citep{likelihood}, in which the morphological and spectral information are fitted simultaneously. Detailed study indicates that a log-parabola function well describes the spectral behavior of the Crab Nebula \citep{LHAASO:science}. The function form assumed for the forward-folded fit is:

\begin{center}
\begin{equation}
\frac{dN}{dE}=\phi_{0}E^{-(a+blog_{10}(E))}\label{function2}.
\end{equation}
\end{center}
The spectral parameters $\phi_{0}$, $a$, $b$, as well as the positional parameters $\sigma$, RA, Dec are selected to maximize the test statistic:
\begin{center}
\begin{equation}
TS\equiv 2\ln\frac{L_{S+B}(\phi_{0},a,b,\sigma,RA,DEC)}{L_{B}}\label{function3},
\end{equation}
\end{center}
$L_{S+B}$ represents the likelihood value for the signal-plus-background hypothesis, while $L_{B}$ represents the value for the background-only hypothesis. The convolution function is obtained by combining the PSF and dN/dE. In this paper, we maximize TS and obtain the six parameters (three from dN/dE and three positional parameters) simultaneously using the Tminuit package in root.
 
\begin{figure}[htbp]
\centering
\includegraphics[width=0.8\textwidth]{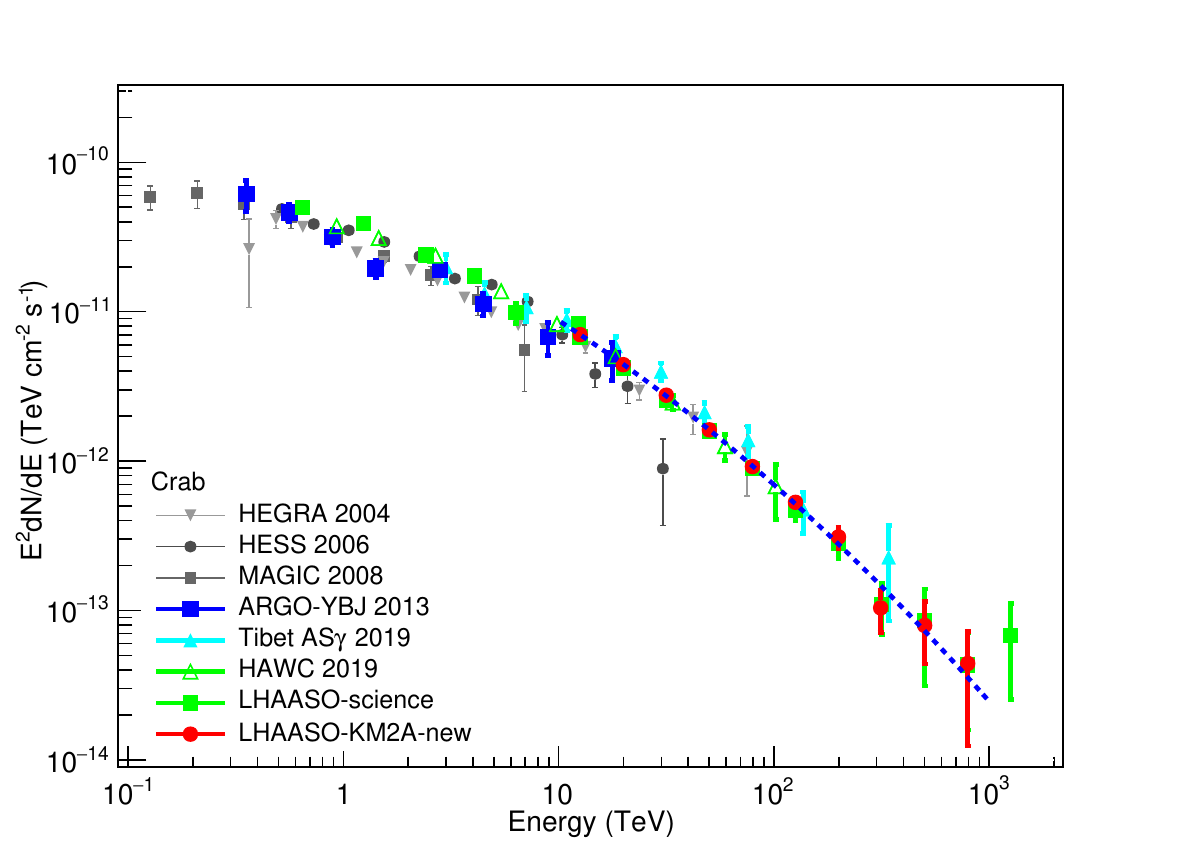}
\caption{Spectrum of the Crab Nebula measured by the KM2A full array using new data selection criteria (red dot). The blue dotted line is the best-fit result using a log-parabola model. The remaining points represent the results of other experiments, including HEGRA \citep{HEGRA}, HESS \citep{HESS_2006}, MAGIC \citep{MAGIC_2008}, ARGO-YBJ \citep{ARGO}, Tibet AS$\gamma$ \citep{ASgamma}, HAWC \citep{HAWC_2019}, and previously published LHAASO results \citep{LHAASO:science}. }
\label{fig:my_label15}
\end{figure}

The inferred SED of the Crab Nebula obtained is shown in Figure \ref{fig:my_label15}. The resulting differential flux ($ \mathrm{TeV}^{-1}\mathrm{cm}^{-2}\mathrm{s}^{-1}$) in the energy range from 10 to 1000 TeV is: 
\begin{center}
\begin{equation}
f(E) = \phi_{0} \left( \frac{E}{10 \text{TeV}} \right)^{-a-\log_{10}(E/10 \text{TeV})}\label{function4},
\end{equation}
\end{center}

where $\phi_{0} = 8.72 \pm 0.10_{\text{stat}}, \quad a = 2.92 \pm 0.04_{\text{stat}}, \quad b = 0.18 \pm 0.04_{\text{stat}}$. For a log-parabola model, the computed TS value amounts to 13464, representing a significant increase of 2334 compared to the old data selection criteria. The SED obtained in this work is generally consistent with several previous experimental results.

\section{Discussion and Conclusions}\label{s:conclusion}
The performance of the KM2A full array for $\gamma$-ray source detection was evaluated using 350 days of data, with the Crab Nebula being analyzed as a ``standard candle". The performance was optimized by adjusting the $\gamma$/hadron discrimination parameter and setting a threshold for $N_{hit}$. Compared with previous studies, the significance of the Crab Nebula was increased by about $20\%$. Additionally, it was found that the pointing error of the KM2A full array is less than 0.03$^{\circ}$, with an angular resolution estimated to be less than 0.2$^{\circ}$ above 100 TeV. The spectrum from 10 TeV to 1000 TeV can be well fitted by a log-parabola model with a spectral index of $(2.92\pm0.04_{\text{stat}})+(0.18\pm0.04_{\text{stat}})\log_{10}(E/10\text{TeV})$, which is consistent with previous measurements by other detectors. 

Our study highlights a significant optimization of significance under 100 TeV, which is beneficial for identifying more low-energy $\gamma$-ray sources such as binaries and quasars. However, sources at large zenith angles are disadvantaged by the new data selection criteria. In the future, the data will be further optimized specifically for observations at large zenith angles.

\begin{acknowledgments}
 We would like to thank all staff members who work at the LHAASO site above 4400 meters above sea level year-round to maintain the detector and keep the water recycling system, electricity power supply and other components of the experiment operating smoothly. We are grateful to the Chengdu Management Committee of Tianfu New Area for the constant financial support for research with LHAASO data. This research work is also supported by the following grants: The National Key R \& D program of China under grants 2018YFA0404201, 2018YFA0404202, 2018YFA0404203, and 2018YFA0404204, by the National Natural Science Foundation of China No.12022502, No.12205314, No.12105301, No.12261160362, No.12105294, No.U1931201, by Youth Innovation Promotion Association CAS(No.2022010) and and in Thailand by the National Science and Technology Development Agency (NSTDA) and National Research Council of Thailand (NRCT): High-Potential Research Team Grant Program (N42A650868).
\end{acknowledgments}

%

\bibliographystyle{aasjournal}
\bibliography{sample631}



\end{document}

%% file: name.tex
\author{Zhen Cao}
\affiliation{Key Laboratory of Particle Astrophyics \& Experimental Physics Division \& Computing Center, Institute of High Energy Physics, Chinese Academy of Sciences, 100049 Beijing, China}
\affiliation{University of Chinese Academy of Sciences, 100049 Beijing, China}
\affiliation{Tianfu Cosmic Ray Research Center, 610000 Chengdu, Sichuan,  China}
 
\author{F. Aharonian}
\affiliation{Dublin Institute for Advanced Studies, 31 Fitzwilliam Place, 2 Dublin, Ireland }
\affiliation{Max-Planck-Institut for Nuclear Physics, P.O. Box 103980, 69029  Heidelberg, Germany}
 
\author{Q. An}
\affiliation{State Key Laboratory of Particle Detection and Electronics, China}
\affiliation{University of Science and Technology of China, 230026 Hefei, Anhui, China}
 
\author{Axikegu}
\affiliation{School of Physical Science and Technology \&  School of Information Science and Technology, Southwest Jiaotong University, 610031 Chengdu, Sichuan, China}
 
\author{Y.X. Bai}
\affiliation{Key Laboratory of Particle Astrophyics \& Experimental Physics Division \& Computing Center, Institute of High Energy Physics, Chinese Academy of Sciences, 100049 Beijing, China}
\affiliation{Tianfu Cosmic Ray Research Center, 610000 Chengdu, Sichuan,  China}
 
\author{Y.W. Bao}
\affiliation{School of Astronomy and Space Science, Nanjing University, 210023 Nanjing, Jiangsu, China}
 
\author{D. Bastieri}
\affiliation{Center for Astrophysics, Guangzhou University, 510006 Guangzhou, Guangdong, China}
 
\author{X.J. Bi}
\affiliation{Key Laboratory of Particle Astrophyics \& Experimental Physics Division \& Computing Center, Institute of High Energy Physics, Chinese Academy of Sciences, 100049 Beijing, China}
\affiliation{University of Chinese Academy of Sciences, 100049 Beijing, China}
\affiliation{Tianfu Cosmic Ray Research Center, 610000 Chengdu, Sichuan,  China}
 
\author{Y.J. Bi}
\affiliation{Key Laboratory of Particle Astrophyics \& Experimental Physics Division \& Computing Center, Institute of High Energy Physics, Chinese Academy of Sciences, 100049 Beijing, China}
\affiliation{Tianfu Cosmic Ray Research Center, 610000 Chengdu, Sichuan,  China}
 
\author{J.T. Cai}
\affiliation{Center for Astrophysics, Guangzhou University, 510006 Guangzhou, Guangdong, China}
 
% ...
\collaboration{300}{The LHAASO Collaboration}